\documentclass[12pt,prd,superscriptaddress,nofootinbib]{article}
\usepackage[english]{babel}
\usepackage{graphicx}
\usepackage{mathtext}
\usepackage{indentfirst}
\usepackage{epsfig,amsmath,amsfonts}

\def\ee{\end{equation}}
\def\ba{\begin{eqnarray}}
\def\ea{\end{eqnarray}}

\def\bq{\begin{quote}}
\def\eq{\end{quote}}

 at 10truept

\newcommand{\beq}{\begin{equation}}
\newcommand{\eeq}{\end{equation}}
\newcommand{\beqa}{\begin{eqnarray}}
\newcommand{\eeqa}{\end{eqnarray}}
\newcommand{\bea}{\begin{eqnarray}}
\newcommand{\eea}{\end{eqnarray}}

 \newcommand{\be}{\beta}

\def\lesssim{~\mbox{\raisebox{-.6ex}{$\stackrel{<}{\sim}$}}~}

\def\ltap{\ \raise.3ex\hbox{$<$\kern-.75em\lower1ex\hbox{$\sim$}}\ }
\def\gtap{\ \raise.3ex\hbox{$>$\kern-.75em\lower1ex\hbox{$\sim$}}\ }
\def\gl{\ \raise.5ex\hbox{$>$}\kern-.8em\lower.5ex\hbox{$<$}\ }
\def\roughly#1{\raise.3ex\hbox{$#1$\kern-.75em\lower1ex\hbox{$\sim$}}}

\newcommand\bqa {\begin{eqnarray}}
\newcommand\eqa {\end{eqnarray}}
\newcommand{\bec}{\begin{cases}}
\newcommand{\eec}{\end{cases}}
\newcommand{\bei}{\begin{itemize}}
\newcommand{\eei}{\end{itemize}}
\newcommand{\bee}{\begin{enumerate}}
\newcommand{\eee}{\end{enumerate}}

\newcommand{\bear}{\begin{array}}
\newcommand{\enar}{\end{array}}

\newcommand{\R}{\mathbb{R}}

\begin{document}

\def\I{{\rm i}}

\def\h{\hbar}

\def\t{\theta}
\def\T{\Theta}
\def\w{\omega}
\def\ov{\overline}
\def\a{\alpha}
\def\b{\beta}
\def\g{\gamma}
\def\s{\sigma}
\def\l{\lambda}
\def\wt{\widetilde}
\def\t{\tilde}


\def \l {\left (}
\def \r {\right )}
\def \R {\mbox{Re}}
\def\be{\begin{eqnarray}}
\def\ee{\end{eqnarray}}


\hfill ITEP--TH--2/13

\vspace{5mm}

\centerline{\Large \bf Infrared dynamics of the massive $\phi^4$ theory on de Sitter space}

\vspace{5mm}

\centerline{E. T. ${\rm Akhmedov}^{1),2),3)}$, F. K. ${\rm Popov}^{1),2)}$ and V. M. ${\rm Slepukhin}^{1),2)}$}


\begin{center}
{\it $\phantom{1}^{1)}$ B. Cheremushkinskaya, 25, Institute for Theoretical and Experimental Physics, 117218, Moscow, Russia}
\end{center}

\begin{center}
{\it $\phantom{1}^{2)}$ Institutskii per, 9, Moscow Institute of Physics and Technology, 141700, Dolgoprudny, Russia}
\end{center}

\begin{center}
{\it $\phantom{1}^{3)}$ Vavilova, 7, Mathematical Faculty of the National Research University Higher School of Economics, 117312, Moscow, Russia}
\end{center}

\vspace{3mm}

\begin{center}{\bf Abstract}\end{center}

We study massive real scalar $\phi^4$ theory in the expanding Poincare patch of de Sitter space.
We calculate the leading two-loop infrared contribution to the two-point function in this theory.  We do that for the massive fields both from the principal and complementary series. As can be expected at this order light fields from the complementary series show stronger infrared effects than the heavy fields from the principal one. For the principal series, unlike the complementary one, we can derive the kinetic equation from the system of Dyson--Schwinger equation, which allows us to sum up the leading infrared contributions from all loops. We find two peculiar solutions of the kinetic equation. One of them describes the stationary Gibbons--Hawking-type distribution for the density per comoving volume. Another solution shows explosive (square root of the pole in finite proper time) growth of the particle number density per comoving volume. That signals the possibility of the destruction of the expanding Poincare patch even by the very massive fields. We conclude with the consideration of the infrared divergences in global de Sitter space and in its contracting Poincare patch.

\vspace{5mm}

\section{Introduction}

There are large infrared (IR) loop contributions even in the massive field theories in the Poincare patch (PP) of de Sitter (dS) space \cite{Krotov:2010ma}, \cite{Jatkar:2011ju}, \cite{Akhmedov:2011pj}. See also \cite{Woodard}--\cite{Xue:2012wi} for the situation in the case of massless fields. In global dS space there are IR divergences \cite{Krotov:2010ma}, \cite{Polyakov:2012uc}, \cite{Akhmedov:2012dn} (see also \cite{Akhmedov:2008pu}), which lead to the inevitable breaking of the dS isometry in the loops for any initial state. They are specific to the quantum fields in dS space and are absent, e.g., in AdS space \cite{Akhmedov:2012hk}.

For the alternative point of view on the IR properties of the massive field theories in dS space, see Refs. \cite{MarolfMorrison}, \cite{Hollands:2010pr}. However, these papers heavily rely on the analytical properties of the correlators as functions of the dS-invariant distances. But such an approach does not work when one wants to understand whether or not the dS-invariant state is stable under small nonsymmetric perturbations above the dS-invariant state because, in the presence of the noninvariant initial density perturbations, even the tree-level two-point correlation functions start to depend on each of their arguments separately rather than on the dS-invariant distances between them. Please note that, in such a case, the vacuum is still dS invariant. We just consider a finite initial particle density over this vacuum. (The notion of the particle is meaningful at least at the past infinity of EPP.)

The reason we consider such density perturbations is that we find it as physically inappropriate to consider the stability of a system in such a state in which all its symmetries are preserved. It is probably worth mentioning here that Minkowski space is stable under the comparatively small noninvariant particle density perturbations over the Poincare-invariant vacuum. That is just a consequence of the energy conservation, which is not present in dS.
Moreover, in global dS, one even does not even have to consider initial density perturbations --- because dS isometry is broken in the loops by the IR divergences.

In Refs. \cite{Krotov:2010ma}, \cite{Jatkar:2011ju}, \cite{Akhmedov:2011pj}, \cite{Polyakov:2012uc}, \cite{Akhmedov:2012dn}, \cite{Akhmedov:2012pa}, the large IR contributions to the two-point functions were calculated for the massive real scalar field theory with the cubic, $\phi^3$, self-interaction. The cubic potential has the runaway instability. To show that those IR effects, which are observed in our previous papers, are universal, we consider here the scalar field theory with $\phi^4$ self-interaction. Furthermore, in our pervious papers, only massive fields from the principal series have been considered. Here, we extend those considerations to the complementary series.

We calculate loop corrections to the so-called Keldysh propagator in the PP of dS space. In the $\phi^4$ theory, there are not any large IR contributions in the first loop. However, at the two-loop order, we find such contributions in the sunset diagrams. In the case of the fields from the principal series, the contribution is linear logarithmic in the physical momentum. For the complementary series, the contribution is powerlike, i.e., it is stronger than for the principal series.

We are not yet able to perform the summation of the higher-loop contributions for the case of the complementary series. But for the principal series, it is possible to do such a summation. That is done via a suitable IR ansatz for the solution of the system of Dyson--Schwinger (DS) equations. This ansatz allows one to reduce this system to a generalization of the Boltzmann kinetic equation. The latter one has a clear physical meaning and describes various particle decay and creation processes in the dS space (see also \cite{Myrhvold}--\cite{Volovik:2008ww}, \cite{Akhmedov:2008pu}). We solve this kinetic equation in two cases. One corresponds to the very mild initial density perturbation over the initial Bunch--Davies state. In such a case, the state of the theory relaxes to the eventual Gibbons--Hawking-type stationary distribution for the density per comoving volume. Another situation corresponds to the strong enough initial density perturbation, which, however, is still much smaller than the cosmological constant. In this case, the state of the theory shows the explosive (square root of the pole in finite proper time) growth of the particle density per comoving volume.

We conclude with the consideration of the contracting PP of dS space and of the global dS and draw similar conclusions to those that have been made in Ref. \cite{Akhmedov:2012dn} for the $\phi^3$ theory.

\section{Setup of the problem}

$D$-dimensional de Sitter (dS) space-time is the hyperboloid, $X_0^2 - X_i^2 = -1$, $i=1,\dots, D$, inside $(D+1)$-dimensional Minkowski space-time, $ds^2 = dX_0^2 - dX_i^2$. Throughout this paper we set its curvature to one and mostly consider its half (e.g. $X_0 \ge X_D$), which is referred to as the expanding Poincare patch (PP): $ds^2 = \frac{1}{\eta^2} (d \eta^2 - d\vec{x}^2)$, where $\eta \in (+\infty, 0)$. Note that while $\eta \to +\infty$ is the past, $\eta\to 0$ corresponds to the future infinity. The action of the theory that we are going to study is

\begin{equation}
S = \int d^D x \, \sqrt{|g|} \, \left[\frac{1}{2} \partial_\mu \phi \partial^\mu \phi  - m^2 \phi^2 - \frac{\lambda}{4!} \phi^4\right].
\end{equation}
Throughout this paper, we always assume that $m > 0$.

Corresponding free harmonics in the Fourier expansion $\phi(x, \eta) = \int d^{D-1}k  \,\eta^{\frac{D-1}{2}} \left[a_k \, h(k \eta)\, e^{-i k x}\right.$ $\left.+ a^+_k \, h^*(k \eta) \, e^{i k x}\right]$, $k = \left|\vec{k}\right|$ are defined via $h(x)$, which is a solution of the Bessel equation with the index $i\, \mu = i\, \sqrt{m^2 - \left(\frac{D-1}{2}\right)^2}$. The choice of such a solution specifies the dS-invariant vacuum $a_k \, |vac\rangle = 0$. For example, the Bunch--Davies (BD) state (or in vacuum in the PP) corresponds to $h(p\eta) = \frac{\sqrt{\pi}}{2} \, e^{-\frac{\pi\mu}{2}}\, H_{i\mu}^{(1)}(p\eta)$, where $H_{i\mu}^{(1)}$ is the Hankel function of the first kind. All other harmonics and vacua can be obtained from those of BD via a one-parameter family of the Bogolyubov rotations \cite{Mottola:1984ar}, \cite{Allen:1985ux}. For example, the out vacuum corresponds to the so-called out Jost harmonics, $h(p\eta) = \sqrt{\frac{\pi}{\sinh (\pi\mu)}}\, J_{i\mu}(p\eta)$, where $J_{i\mu}$ is the Bessel function of the first kind. Conjugate harmonics in the latter case are given by the $Y$'s --- Bessel functions of the second kind.

In the nonstationary situation every particle is described by the matrix propagator (see, e.g., Ref. \cite{vanderMeulen:2007ah} for the Feynman rules in the $\phi^3$ scalar field theory on dS space), for which the entries are the Keldysh propagator, $G_K = \frac12 \, \langle \left\{\phi(x), \phi(y)\right\} \rangle$,
and the retarded and advanced propagators\footnote{Here, $\Delta x_0 = x_0 - y_0$, $\{,\}$ is the anticommutator and $[,]$ is the commutator; $\theta(\eta)$ is the Heaviside $\theta$ function.}, $G^A_R = \mp \langle \left[\phi (x), \phi(y)\right] \rangle\, \theta(\mp\Delta x_0)$. The more detailed discussion of the physics in dS space, which is relevant for the present paper, can be found in Refs. \cite{Krotov:2010ma}, \cite{Akhmedov:2011pj}, \cite{Polyakov:2012uc}, \cite{Akhmedov:2012dn}, \cite{Akhmedov:2012pa}. It is instructive for the further discussion to keep in mind that the result of the calculation with the use of the nonstationary Keldysh--Schwinger diagrammatic technique provides a solution to a sort of the Cauchy problem.

Because the PP is spatially homogeneous and because below we concentrate on the spatially homogeneous states, we find it convenient to make the spatial Fourier transform of the aforementioned propagators:

\be\label{oneloop1}
D^{K,R,A}\left(\eta_1, \eta_2, p\right) = \int d^{D-1} x \, e^{i \vec{p}\, \vec{x}}\, G^{K,R,A} \left(\eta_1, \vec{x}, \eta_2, 0\right), \nonumber \\
D^K(\eta_1, \eta_2,p) = (\eta_1 \eta_2)^{\frac{D - 1}{2}} d^K(p \eta_1, p \eta_2), \nonumber \\
D^A_R(\eta_1, \eta_2,p)=\mp\theta(\pm \Delta\eta)(\eta_1 \eta_2)^{\frac{D-1}{2}} d^-(p\eta_1,p\eta_2),
\ee
where $\Delta \eta = \eta_1 - \eta_2$, $p = \left|\vec{p}\right|$ and

\be\label{oneloop}
d^-(p\eta_1,p\eta_2)=2 \mbox{Im}\left[h(p\eta_1)h^*(p\eta_2)\right], \nonumber \\
d^K(p\eta_1,p\eta_2)  =  h(p\eta_1) h^*(p \eta_2) \l \frac{1}{2} + n_p \r +h(p\eta_1)h(p\eta_2)k_p+\mbox{c.c.}\, .
\ee
Here, we use the convenient notations $n_p = \langle a^+_p\, a_p \rangle$ and $\kappa_p = \langle a_p \, a_{-p}\rangle$. In the nonstationary situation, $n_p$ and $k_p$ can be zero only on the tree level, i.e., if the initial state is chosen to be $|vac\rangle$.

It is worth stressing here that, in our paper, we always study the two-point correlation function. According to Eqs. (\ref{oneloop1})--(\ref{oneloop}), the quantities $n_p = \langle a^+_p \, a_p \rangle$ and $k_p = \langle a_p \, a_{-p}\rangle$ are just elements of this correlation function. Hence, all our conclusions about their behavior have invariant physical meaning. Furthermore, in the situation in which the anomalous quantum average $k_p$ vanishes, $n_p$ acquires the clear physical meaning --- it becomes the particle number density per comoving volume in the given state of the theory. In fact, then free Hamiltonian of the theory acquires the diagonal form. Moreover, then all elements of the kinetic equation that is presented below acquire the clear physical meaning.

Let us provide here a few more arguments favoring our interpretation of $n_p$ as the particle density.
Various observers may indeed detect different particle fluxes. However, one should separate the Unruh effect from what we would like to call the real particle production. In Minkowski space, both inertial and noninertial observers see the same state --- Minkowski (Poincare invariant) vacuum. However, while the inertial observer sees it as the empty space, the noninertial one sees it as the thermal state. That is due to the specific correlation of the vacuum fluctuations along its worldline \cite{Akhmedov:2007xu}. Note that there is not any nontrivial gravitational field in the circumstances under consideration because the Riemanian tensor is exactly zero.

The real particle creation is due to the change of the ``vacuum'' state under the influence of the background field. That is exactly what happens in the background electric field in dS space and in the collapsing black hole background.

Rephrasing, we would like to say here that, while in Minkowski space there is one type of observers that does not see any particle flux, in dS space, there is no such observer that sees nothing. On general grounds, we expect that the least particle flux is seen by inertial observers --- they do not see the extra Unruh type of flux, so to say. Our calculations are actually done for the inertial observers because the change from the proper time $t$ to the conformal one, $\eta = e^{-t}$, is just the change of the clock's rate rather than a transition to some nontrivial motion. In any case, all our formulas can be trivially rewritten in the proper time $t$.

Finally, the IR dynamics depends on the value of the mass $m$ and on the choice of $h(x)$. Below, we separately consider the following two cases:  $\left(\frac{D-1}{2}\right)^2 < m^2$ (the principal series) and $\left(\frac{D-1}{2}\right)^2 > m^2$ (the complementary series). The crucial physical difference between these two cases is due to the fact that, while the harmonics of the principal series oscillate at the future infinity, $h(x) \approx A_+ \, x^{i\mu} + A_- \, x^{-i\mu}$, $x\to 0$, those of the complementary series do not do that because, in the latter case, $\mu$ is pure imaginary.

\section{Two-loop contribution}

In this section, we calculate loop corrections to the Keldysh propagator, $D^K(\eta_1,\eta_2,p)$, with the initial  dS invariant vacuum state at the past infinity of the PP. In these settings the tree-level $D^K(\eta_1, \eta_2, p)$ is given by Eq. (\ref{oneloop}) with $n_p=0$ and $\kappa_p = 0$. It is straightforward to show that the $\phi^4$ theory, unlike the $\phi^3$ one, does not possess any large IR contributions to any propagator at the first-loop order ($\sim \lambda$). However, in the second-loop order ($\sim \lambda^2$), there is a large IR contribution to $D^K$, which is of interest for us. We consider the IR limit in which $p\sqrt{\eta_1\, \eta_2} \to 0$ and $\eta_1/\eta_2 = const$ \cite{Akhmedov:2011pj}, \cite{Akhmedov:2012pa}, \cite{Akhmedov:2012dn}. It corresponds to the situation in which both time arguments of the propagator are taken to the future infinity, while the time distance between them is kept finite.

The reason we pay attention only to the Keldysh propagator is that it defines the state of the theory, i.e., shows the dependence of $n_p$ and $k_p$ on time. Moreover, in the case of the principal series, it is the only propagator that receives large corrections of the order $\lambda^2 \log (p\,\sqrt{\eta_1\eta_2}/\mu)$ in the IR limit in question. In fact, in Ref. \cite{Jatkar:2011ju}, it was shown that the retarded and advanced propagators of the $\phi^3$ theory receive only finite corrections from the first loop --- of the order $\lambda^2 \log(\eta_1/\eta_2)$. In the case of the $\phi^4$ theory, the situation is similar in the second loop. Furthermore, it is straightforward to show that, for the principal series, the large IR contributions to the interaction vertex are also suppressed by higher powers of $\lambda$. For the complementary series, the situation is more subtle, but we still would like to start their study with the consideration of the loop corrections to $D^K$.

The large IR loop contribution to $D^K$ (if there is one) at the order $\lambda^2$ can be expressed in the form (\ref{oneloop1}), (\ref{oneloop}) with

\be\label{6}
n_p( \eta) \approx - \frac{\lambda^2}{3\, (2\pi)^{2(D-1)}} \int d^{D-1} q_1 \, d^{D-1} q_2 \, d^{D-1} q_3 \, \int^\eta_\infty d \eta_3 \, \int^\eta_\infty d \eta_4 \, (\eta_3 \eta_4)^{D-2} \,
\nonumber \\ \delta^{(D-1)} \left(\vec{p} + \vec{q}_1 + \vec{q}_2 + \vec{q}_3\right)
h(p\eta_3) h(q_1\eta_3) h(q_2 \eta_3) h(q_3 \eta_3) h^*(p\eta_4)
h^*(q_1\eta_4) h^*(q_2 \eta_4) h^*(q_3 \eta_4),
\nonumber \\
k_p( \eta) \approx \frac{2\, \lambda^2}{3\, (2\pi)^{2(D-1)}}\, \int d^{D-1} q_1 \, d^{D-1} q_2 \, d^{D-1} q_3 \, \int_\infty^\eta d \eta_3 \int_\infty^{\eta_3} d \eta_4 (\eta_3 \eta_4)^{D-2}
\nonumber \\ \delta^{(D-1)} \left(\vec{p} + \vec{q}_1 + \vec{q}_2 + \vec{q}_3\right)
h^*(p\eta_3) h(q_1\eta_3) h(q_2 \eta_3) h(q_3 \eta_3) h^*(p\eta_4)
h^*(q_1\eta_4) h^*(q_2 \eta_4) h^*(q_3\eta_4).
\ee
Here, $p = \left|\vec{p}\right|$, $q_{1,2,3} = \left|\vec{q}_{1,2,3}\right|$, and $n_p(\eta)$ is real. Note that, if one will take the flat space limit of these expressions, i.e., substitute $\eta$ by $t$, $\sqrt{|g|}$ by 1, and $\eta^{(D-1)/2}\, h(k\eta)$ by $e^{i \, \epsilon(k) \, t}$, then $n_p$ and $k_p$ would vanish, when $\eta\to 0$, as the consequence of the energy conservation.

In deriving these expressions, we have used that $h(p\eta)$ depends only on $\left|\vec{p}\right|$,
and, hence, we can safely change $\vec{p} \rightarrow - \vec{p}$. Also in the IR limit in question inside the leading loop corrections, one can neglect the difference between $\eta_1$, $\eta_2$, and $\eta = \sqrt{\eta_1\, \eta_2}$. In such an approximation, we drop the subleading, $\lambda^2 \log\left(\eta_1/\eta_2\right)$, contributions from the expressions for $n_p$ and $k_p$. The derivation of Eq. (\ref{6}) is similar to the one performed in Refs. \cite{Krotov:2010ma}, \cite{Jatkar:2011ju}, \cite{Akhmedov:2011pj}, \cite{Akhmedov:2012dn}, \cite{Akhmedov:2012pa} for the $\phi^3$ theory.

\subsection{Principal series}

We start with the case $ \frac{D-1}{2} < m $ and make the following change of the integration variables in Eq. (\ref{6}): $\vec{q}_i$ to $\vec{l}_i = \vec{q}_i \eta_3$ and $\eta_4$ to $v = \frac{\eta_3}{\eta_4}$, $i=1,2,3$. Then, we expand $h(p\eta_{3,4}) \approx A_+ (p\eta_{3,4})^{i\mu} + A_- (p\eta_{3,4})^{-i\mu}$ as $p\eta_{3,4} \to 0$ under the integrals, where $A_{\pm}$ are some mass-dependent complex constants. After that, we neglect\footnote{The justification of all the approximations listed here can be found in Refs. \cite{Krotov:2010ma}, \cite{Jatkar:2011ju}, \cite{Akhmedov:2011pj}, \cite{Polyakov:2012uc}, \cite{Akhmedov:2012pa}, \cite{Akhmedov:2012dn}. In particular, the reason we neglect $p$ in comparison with $q_{1,2,3}$ is that the largest IR contribution comes from this region of integration over $q_{1,2,3}$.} $p$ in comparison with $q_{1,2,3}$ on the RHS of Eq. (\ref{6}) and perform the integration over $\eta_3$. The largest IR contributions come from the integrals of the type $\int_{\mu/p}^\eta \frac{d \eta_3}{\eta_3} h\left(p \, \eta_3\right) h^* \left(p\, \eta_3 \, v\right)$
and $\int_{\mu/p}^\eta \frac{d \eta_3}{\eta_3} h^*\left(p \, \eta_3\right) \, h^*\left(p\,\eta_3 \, v\right)$, where $h$'s are Taylor expanded. The result is as follows:

\be\label{nkappa}
n_{p\eta} \approx - \frac{\lambda^2\, \log(p \eta/\mu)}{3\,(2\pi)^{2(D-1)}} \int_{\infty}^{0} dv v^{D-2} \, \int d^{D-1} l_1 \, d^{D-1} l_2 \, d^{D-1} l_3 \, \delta^{(D-1)} \l \vec{l}_1 + \vec{l}_2 +  \vec{l}_3 \r \times \nonumber \\ \times
h^*(l_1) h \l l_1 v \r
h^*(l_2) h \l l_2 v \r
h^*(l_3) h \l l_3 v \r \left[\left|A_+\right|^2\, v^{-i\mu} + \left|A_-\right|^2\, v^{+i\mu}\right], \nonumber \\
k_{p\eta} \approx \frac{2\lambda^2 \, \log(p \eta/\mu)}{3\,(2\pi)^{2(D-1)}} \int_{\infty}^{1} dv v^{D-2} \, \int d^{D-1} l_1 \, d^{D-1} l_2  \, d^{D-1} l_3 \, \delta^{(D-1)} \l \vec{l}_1 + \vec{l}_2 +  \vec{l}_3  \r \times \nonumber \\ \times h^*(l_1) h \l l_1 v \r
h^*(l_2) h \l l_2 v \r
h^*(l_3) h \l l_3 v \r A_+ A_- \left[v^{i\mu} + v^{-i\mu}\right], \quad l_{1,2,3} = \left|\vec{l}_{1,2,3}\right|.
\ee
The lower limit of integration over $\eta_3$ we cut by $\mu$ because, at $p\eta \gg \mu$ the $d^{D-1}l_i$, $d\eta_3$, and $dv$, integrals are rapidly convergent due to the oscillations of $h(x)$, while we care only about the leading IR contribution. It is worth stressing here that such a convergence of the $dv$ and $d^{D-1} l_i$ integrals is true in the sense of the generalized functions. The latter fact is related to the behavior of the modes with high momenta, discussed below.

Note that $n_p(\eta)$ and $k_p(\eta)$ are functions of the physical momentum, $p\eta$, rather than separately depending on the momentum, $p$, and on the time, $\eta$. That is natural because of the spatial homogeneity and the invariance of the PP dS metric under the simultaneous rescaling $\vec{x} \to \sigma \vec{x}$, $\eta \to \sigma \eta$.

It was explained in Refs. \cite{Akhmedov:2011pj}, \cite{Polyakov:2012uc}, \cite{Akhmedov:2012dn}, \cite{Akhmedov:2012pa} that such large log contributions in $D^K$ appear due to the particle creation in dS space (see also the discussion below). In the expanding PP, the creation of particles with comoving momentum $p$ effectively starts after some moment of time $\eta_* \lesssim \mu/p$ because the modes with high momenta, $p\eta \gg \mu$, do not feel the curvature of the space-time and behave as if they are in flat space. As a result, $n_p$ and $k_p$ are proportional to the proper time, $\log\left(\mu/p\eta\right)$, elapsed from $\mu/p$ to $\eta$. The coefficient of the proportionality is just the particle production rate. It is worth stressing here that the presence of the large $k_p$, which is comparable to $n_p$, signals that there is the strong backreaction on the initial state $|vac\rangle$ \cite{Akhmedov:2011pj},\cite{Akhmedov:2012dn},\cite{Akhmedov:2012pa}.

For further reference, it is instructive to study also the IR behavior of $n_p$ and $k_p$ for the out Jost harmonics.
Although the UV behavior of the correlation functions for the out Jost harmonics is different from the proper flat space type, we are interested here in the IR limit, in which out Jost harmonics may play a crucial role\footnote{While the theory under consideration shows the proper UV behavior in the BD state, it does not do that in any other $\alpha$ state. The reason for that is as follows. While BD harmonics behave as single waves $e^{i \, p \, \eta}$ in the UV limit, $p\eta \to \infty$, the other $\alpha$ harmonics behave as linear combinations of $e^{i p \eta}$ and $e^{-i p \eta}$ in the same limit.}. In fact, in condensed matter physics, it is the frequent situation that, in the IR limit, one has to perform a Bogolyubov rotation to some harmonics for which the UV properties may be different from the proper ones; however, they properly describe the IR physics. The seminal example is the BCS theory for superconductivity.

In particular, we will see that the proper solution of the IR limit of the system of DS equations is obtained via the out Jost harmonics. The hint for that comes from the following observation.
For the out Jost harmonics, the leading IR two-loop contribution has a crucial difference with respect to that of BD harmonics or any other $\alpha$ harmonics. In fact, the out Jost harmonics behave as $h(p\eta_{3,4}) \approx A (p\eta_{3,4})^{i\mu}$ in the future infinity, $p\eta_{3,4}\to 0$, where $A$ is some mass-dependent complex constant. Then, it is straightforward to show that $n_p$ has the same form as Eq. (\ref{nkappa}) with $|A|^2 \,  v^{i \mu}$ instead of $\left[\left|A_+\right|^2 \, v^{i\mu} + \left|A_-\right|^2\, v^{-i\mu}\right]$. At the same time, $k_p$ does not receive any large contributions in the IR limit in question; i.e., it is negligible in comparison with $n_p$. This is going to be an important observation for the derivation of the kinetic equation below.

It is worth stressing here that, for the out Jost harmonics, the $dv$ and $d^{D-1}l_i$ integrals in the obtained expressions for $n_p$ and $k_p$ are also convergent. In fact, the situation in the $\phi^4$ theory is even simpler than in the $\phi^3$ case \cite{Akhmedov:2011pj}.

\subsection{Complementary series}

We continue with the consideration of the complementary series, $\frac{D-1}{2} > m$, corresponding to the imaginary $\mu$, i.e. to the real index of the solution of the Bessel equation $h(p\eta)$. Below, we use the notation $\nu = - i\mu$. Then, for the in harmonics, $h = J_\nu + iY_\nu$, where both Bessel functions $J_\nu$ and $Y_\nu$ are real in the case of the real index $\nu$. Expanding them near zero, we get $Y_\nu(x) \approx A_- x^{-\nu} + B x^{ -\nu + 2}$ and $J_\nu(x) \approx A_+ x^{\nu}$.
Because of the possible differences between the behavior of $h$ and $h^*$ near zero, we have to pay attention separately to $k_p$ and $k_p^*$.

The contributions to $n_p$ and $k_p$, $k^*_p$ can be expressed as

\be\label{66}
n_p(\eta) \approx - \frac{2 \lambda^2}{3\,(2\,\pi)^{2\, (D-1)}} \, \int^0_\infty \frac{du}{u} \, \int^0_\infty \frac{dv}{v^{2D - 1}} \, F[v] \, h(u v)  h^*\left(\frac{u}{v}\right) \,
\theta[uv - p \eta] \, \theta\left[\frac{u}{v}  - p \eta\right],
\nonumber \\
k_p(\eta) \approx \frac{4\, \lambda^2}{3\,(2\pi)^{2(D-1)}} \, \int^0_\infty \frac{du}{u} \int_\infty^0 \frac{dv}{v^{2D - 1}} \, F[v] \, h^*(u v)  h^*\left(\frac{u}{v}\right) \, \theta[uv - p \eta] \, \theta\left[\frac{u}{v}  -  u v\right],
\nonumber \\
k^*_p(\eta) \approx \frac{4\, \lambda^2}{3\,(2\pi)^{2(D-1)}} \, \int^0_\infty \frac{du}{u} \int_\infty^0 \frac{dv}{v^{2D - 1}} \, F[v] \, h(u v)  h\left(\frac{u}{v}\right) \, \theta[uv - p \eta] \, \theta\left[\frac{u}{v}  -  u v\right],
\ee
with the use of the following notations

\be
F(\eta_3, \eta_4) = \int \left[\prod_{i=1}^3 d^{D-1} q_i h(q_i\eta_3) h^*(q_i \eta_4)\right]\, (\eta_3 \eta_4)^{D-2} \, \delta^{(D-1)}\left(\vec{p} - \vec{q}_1 - \vec{q}_2 - \vec{q}_3\right),
\ee
which, after the change of integration variables $u=p \sqrt{\eta_3 \eta_4}$, $v=\sqrt{\frac{\eta_3}{\eta_4}}$,
can be expressed as $F(\eta_3, \eta_4) = \frac{p^2}{(uv)^{2 \, (D - 1)}} \, u^{2 \, (D - 2)} \, F[v]$, where $F[v] = F^*\left[1/v\right]$ is some function of one variable $v$.

The leading correction to $n_p$ and $k_p$ is given by Eq. (\ref{66}), where from the Hankel functions, $h(uv)$ and $h(u/v)$, we single out only $Y$'s. Such a contribution gives for $n_p$ and $k_p$ the inverse powerlike behavior in $p\eta$, which, however, cancels out after the substitution into $D^K$ because $Y$ is real. The next order
is obtained as follows. One also has to express $h(p\eta_{1,2})$ through $J_\nu$ and $Y_\nu$ in the full propagator $D^K$. Then, from one of the four $h$'s $[h(p\eta_{1,2})$ and $h(uv)$, $h(u/v)]$, we have to single out $J_\nu$, while from the other three --- $Y_\nu$'s.
This expression does not cancel out and provides the leading IR contribution to $D^K$:

\be
D^K(\eta_1, \eta_2, p) = \frac{8\,\lambda^2 \, A_-^3 \, A_+ \, \eta^{D-1}}{3\,(2\pi)^{2\, (D-1)}} \,
\int \limits_\infty^0 du \, u^{-1-2 \nu} \int \limits_\infty^0 dv \, v^{1-2\,D} F[v] \nonumber \\
\left\{\left[-1+\left(\frac{u}{p\eta \, v}\right)^{2 \nu}\right] \theta \left[uv - \frac{u}{v}\right]\, \theta\left[-p\eta + uv\right] + \left[1-\left(\frac{u
v}{p\eta}\right)^{2 \nu }\right] \, \theta\left[- u v + \frac{u}{v}\right] \, \theta\left[-p\eta + \frac{u}{v}\right] \right\}.
\ee
After the straightforward manipulations, the obtained expression can be reduced to

\be\label{9}
D^K(\eta_1,\eta_2, p) = - \frac{8\,\lambda^2 \, A_-^3\,A_+ \,\eta^{D-1} \, \log(p\eta)}{3\,(2\pi)^{2\, (D-1)} \, (p\eta)^{2 \nu}}\, \nonumber \\
\left\{\int \limits^\infty_1 dv \, v^{1 - 2D}\, F[v] \, \left[-\frac{1}{2 \nu} v^{2 \nu} + \left(\frac{1}{v}\right)^{2 \nu}\right] - \int \limits^1_0 dv \, v^{1 - 2D} \, F[v]\,  \left[\frac{1}{2 \nu} \, ( v)^{-2 \nu} +
v^{2 \nu}\right]\right\}.
\ee
The integral over $v$ is convergent in the IR limit (as $v\to \infty$) if $D > 1 + 4 \nu$. In the UV limit $(v\to 0)$, it is convergent in the sense of the generalized function.

For the out harmonics, the situation is a bit different. In this case, $h = J_\nu$, $h^* = Y_\nu$. The straightforward calculation shows that

\be
n_p(\eta) \propto \lambda^2 \, A_- \, A_+ \, \log(p \eta) \, \int_0^1 dv \, F[v] v^{2\nu + 1 - 2D},
\nonumber \\
k_p(\eta) \propto \lambda^2 \, A_-^2 \, (p \eta)^{-2 \nu} \, \int^0_1 dv \, F[v] \, v^{2\nu + 1 - 2D}, \nonumber \\
k^*_p(\eta) \propto \lambda^2 \, A_+^2 \, (p \eta)^{2 \, \nu} \int_0^1 dv \, F[v] \, v^{-2\nu + 1 - 2D}.
\ee
After the substitution into the Keldysh propagator, the leading contribution comes from $n_p$
and is as follows:

\be
D^K \approx \frac{\lambda^2 \, A_-^2 \, A_+^2 \, \log(p\eta)}{3\, (2\pi)^{D-1}} \, \eta^{D-1} \, \int_0^1 d v   \, v^{2 \nu + 1 - 2D}  F[v]
\ee
Because of the character of these IR contributions for the light fields, we do not yet understand their physical meaning and think that the  kinetic equation obtained in the next section is not applicable for the fields from the complementary series. In the case of the complementary series, we do not yet know how to perform the summation of the leading IR contributions from all loops. But on general physical grounds and from the two-loop result, we expect that complementary series will show stronger IR effects than the heavy fields from the principal series, which is under study below.

\section{ Kinetic equation}

Although $\lambda^2$ is small, the product $\lambda^2 \log \left(\mu/p\eta\right)$ can become large as $p\eta \to 0$. Hence, higher loops are not suppressed in comparison with the tree-level contribution. Then, one has to perform the summation of the leading IR contributions from all loops. This can be achieved via an approximate IR solution of the system of DS equations in the nonstationary diagrammatic technique. We can do that only for the case of the principal series, $m>(D-1)/2$, at least because, in this case, harmonics oscillate in the IR limit. As the result, there is a clear separation of scales between the time dependence of the harmonics $h(p\eta)$ and that of $n_{p\eta}$ and $k_{p\eta}$. This allows us to simplify DS equations in the limit under consideration and eventually to solve them approximately. In the case of the complementary series, however, the character of the IR contributions to $D^K$ and to the vertex does not allow us to have a clear kinetic interpretation.

Putting it in other words, to move further, it is worth observing that, above,
we have calculated the loop corrections to the Keldysh propagator under the assumption that $n_p$ and $k_p$ retain their initial values throughout all the time evolution. In fact, in the calculations of the previous section, we have used tree-level propagators, i.e., their initial values. To make the problem self-consistent, one has to take into account the change of $n_p$ and $k_p$ in time. As we will see, this also allows us to reduce the problem to the solution of the DS equations for the nonstationary diagrammatic technique. The latter ones represent a system of equations for the matrix propagators, self-energy, and vertex. In some circumstances, as we will see, this system can be simplified and reduced to a single equation.

We assume that the evolution had started with some density perturbation over the BD state at the past infinity of the PP. Then, the UV behavior of the theory in question is the same as in flat space, but the dS invariance is slightly broken. We would like to trace the destiny of these density perturbations in the future infinity, i.e., would like to see whether the theory relaxes back to the dS-invariant state or this density explodes, causing the modification of the background geometry. The answer to the latter question also can be obtained only after the solution of the IR limit of the nonstationary DS equation.

Thus, we would like to sum up leading contributions, which are powers of $\lambda^2 \log (p\sqrt{\eta_1\eta_2}/\mu)$, and drop the subleading terms, which are suppressed by higher powers of $\lambda$ and/or powers of $\lambda^2 \log(\eta_1/\eta_2)$. Having in mind that retarded and advanced propagators and the vertex receive only subleading IR contributions in the limit in question\footnote{There are large IR contributions to $D^{A,R}$ and to the vertex, which are coming from those in $D^K$. But they are suppressed by the higher powers of $\lambda$ because they appear in the higher loops. Actually all one-loop contributions to the vertex have similar structure to the one shown in Eq. (\ref{6}) with one momentum integration less. It is straightforward to see that, because of that, there is not any large IR contribution to the vertex at the $\lambda^2$ order.}, we can assume that they take their classical (UV renormalized) values. At the same time we assume that the ansatz for the exact Keldysh propagator is given by Eqs. (\ref{oneloop1}),(\ref{oneloop}) with undefined $n_p(\eta)$ and $k_p(\eta)$, where $\eta = \sqrt{\eta_1\eta_2}$.

Then, the relevant part of the system of DS equations has the form

\be\label{DS}
D^K(\eta_1,\eta_2,p) = D^K_0(\eta_1,\eta_2,p) - \nonumber \\ - \frac{\lambda^2}{6 \, (2\pi)^{2(D-1)}}\, \int d^{D-1} q_1 \, d^{D-1} q_2 \, d^{D-1} q_3 \iint_\infty^0 \frac{d\eta_3 d\eta_4}{(\eta_3 \eta_4)^D} \delta^{(D-1)} \left(\vec{p} - \vec{q}_1 - \vec{q}_2 - \vec{q}_3\right) \nonumber \\
\Bigg[3\, D^K_0(\eta_1,\eta_3,p) D^K(\eta_3,\eta_4,q_1) D^K(\eta_3,\eta_4,q_2) D_0^A(\eta_3,\eta_4,q_3)
D_0^A( \eta_4,\eta_2,p)\nonumber \\
-\frac{1}{4} D^K_0(\eta_1,\eta_3,p) D_0^A(\eta_3,\eta_4,q_1) D_0^A(\eta_3,\eta_4,q_2) D_0^A(\eta_3, \eta_4,q_3) D_0^A( \eta_4,\eta_2,p)\nonumber \\ - \frac{3}{4} \, D^R_0(\eta_1,\eta_3,p) D^K(\eta_3,\eta_4,q_1)
D_0^A(\eta_3, \eta_4,q_2) D_0^A(\eta_3,\eta_4,q_3) D_0^A( \eta_4,\eta_2,p)\nonumber \\
+ D^R_0(\eta_1,\eta_3,p) D^K(\eta_3,\eta_4,q_1) D^K(\eta_3,\eta_4,q_2) D^K(\eta_3,\eta_4,q_3) D_0^A(\eta_4,\eta_2,p)\nonumber \\
-   \frac{3}{4}\, D^R_0(\eta_1,\eta_3,p) D^K(\eta_3,\eta_4,q_1) D_0^R(\eta_3, \eta_4,q_2) D_0^R(\eta_3,\eta_4,q_3)
D_0^A(\eta_4,\eta_2,p) \nonumber \\
- \frac{1}{4} \, D^R_0(\eta_1,\eta_3,p) D_0^R(\eta_3,\eta_4,q_1) D_0^R(\eta_3,\eta_4,q_2) D_0^R(\eta_3, \eta_4, q_3)D^K(p \eta_4,p\eta_2) \nonumber \\ + 3\, D^R_0(\eta_1,\eta_3,p) D_0^R(\eta_3, \eta_4,q_1) D^K(\eta_3, \eta_4,q_2) D^K(\eta_3, \eta_4, q_3) D^K(\eta_4,\eta_2, p)\Bigg] \label{main}
\ee
where $p=\left|\vec{p}\right|$, $q_{1,2,3} = \left|\vec{q}_{1,2,3}\right|$ and $D^{A,R,K}_0$ are the Fourier transforms of the initial values of the retarded, advanced, and Keldysh propagators\footnote{I.e., $D^K_0$ is also given by Eqs. (\ref{oneloop1}),(\ref{oneloop}) with some initial values $n^{(0)}_p$ and $k_p^{(0)}$.}; $D^K$ is the exact Keldysh propagator. This equation is covariant under the Bogolyubov rotations between different harmonics, $h(x)$.
Because we are interested in its solution in the IR limit, we do not have to care about the proper UV behavior and have to check the situation for all possible $\alpha$ harmonics.

We would like to pick out the largest IR contribution from the integral on the RHS of Eq. (\ref{DS}). The calculation is just a straightforward generalization of the above two-loop one. Similarly to Refs. \cite{Akhmedov:2011pj}, \cite{Akhmedov:2012dn}, \cite{Akhmedov:2012pa}, for all $\alpha$ harmonics, including BD ones, we obtain that the ansatz (\ref{oneloop1}), (\ref{oneloop}) solves DS equation in the IR limit in question. But $k_p$ is comparable to $n_p$. This means that the backreaction on the background state (specified by the choice of the harmonics) is big. The only exception is given by the out Jost harmonics. For these harmonics, $k_p$ remains zero if its initial value was zero. Moreover, small perturbations of $k_p$ relax back to zero. For the $\phi^3$, theory this was shown in Ref. \cite{Akhmedov:2012pa}. For the $\phi^4$ theory, the situation is similar.

As the result, with the use of the out Jost harmonics, the IR limit of the DS equation is solved by Eqs. (\ref{oneloop1}), (\ref{oneloop}) with $k_p=0$. We would like to convert the integral DS equation (\ref{DS}) into the integrodifferential form, i.e., into the form of the kinetic equation \cite{LL}. This is done via a kind of the renormalization group procedure as follows \cite{Akhmedov:2012pa}. In the given settings, $n^{(0)}_p$ is the particle density at some moment after $\eta_* \sim \mu/p$. In fact, as we have mentioned above and will explain in the next section, before this moment, all the kinetic processes, except maybe the irrelevant scattering one, are suppressed. Hence, $n_p(\eta)$ remains constant and is equal to $n^{(0)}_p$. Then, from Eq. (\ref{DS}), it is straightforward to derive that the difference between $n_p(\eta)$ and $n_p^{(0)} = n_p(\eta_*)$ is proportional to the proper time elapsed from $\eta_*$ to $\eta$. The coefficient of the proportionality is the collision integral --- the RHS of the kinetic equation is

\be\label{kineq}
\frac{n_p(\eta) - n_p(\eta_*)}{\log(\eta) - \log(\eta_*)} \to \frac{d n_{p\eta}}{d \log(p\eta)} = - \frac{\lambda^2 \, |A|^2}{6} \, \int \frac{d^{D-1} l_1}{(2\pi)^{D-1}} \frac{d^{D-1} l_2}{(2 \pi)^{D-1}} \int_\infty^0 dv \, v^{D-2}\, \nonumber \\
\left\{3 \R \left [v^{i\mu\phantom{\frac{a}{b}}} \,h^*(l_1) h^*(l_2) h\left(\left|\vec{l}_1 + \vec{l}_2\right|\right) h(l_1 v) h(l_2 v) h^*\left(\left|\vec{l}_1 + \vec{l}_2\right| v\right) \right   ] \right. \times \nonumber \\ \times \left[(1+n_{p\eta}) n_{l_1} n_{l_2} (1+n_{|\vec{l}_1 + \vec{l}_2|}) \,\, - \phantom{\frac{a}{b}} n_{p\eta} (1+n_{l_1}) (1+n_{l_2}) n_{|\vec{l}_1 + \vec{l}_2|} \right]
\nonumber \\+ 3  \R \left [v^{i\mu\phantom{\frac{a}{b}}} \,h^*(l_1) h(l_2) h\left(\left|\vec{l}_1 - \vec{l}_2\right|\right) h(l_1 v) h^*(l_2 v) h^*\left(\left|\vec{l}_1 - \vec{l}_2\right| v\right) \right] \times \nonumber \\ \times \left  [(1+n_{p\eta}) n_{l_1} (1+n_{l_2}) (1+n_{|\vec{l}_1 - \vec{l}_2|}) \,\, - \phantom{\frac{a}{b}} n_{p\eta} (1+n_{l_1}) n_{l_2} n_{|\vec{l}_1 - \vec{l}_2|}\right ] \nonumber \\ +
 \R \left [v^{i\mu \phantom{\frac{a}{b}}} \, h^*(l_1) h^*(l_2) h^*\left(\left|\vec{l}_1 + \vec{l}_2\right|\right) h(l_1 v) h(l_2 v) h\left(\left|\vec{l}_1+\vec{l}_2\right| v\right) \right   ]\times \nonumber \\ \times \left  [(1+n_{p\eta}) n_{l_1} n_{l_2} n_{|\vec{l}_1 + \vec{l}_2|} \,\, -  \phantom{\frac{a}{b}} n_{p\eta} (1+n_{l_1}) (1+n_{l_2}) (1+n_{|\vec{l}_1 + \vec{l}_2|})  \right ] \nonumber \\ +
 \R  \left [v^{i\mu \phantom{\frac{a}{b}}} \, h(l_1) h(l_2) h\left(\left|\vec{l}_1 + \vec{l}_2\right|\right) h^*(l_1 v) h^*(l_2 v) h^*\left(\left|\vec{l}_1+\vec{l}_2\right| v\right) \right   ]\times \nonumber \\ \times \left. \left  [ (1+n_{p\eta}) (1+n_{l_1}) (1+n_{l_2}) (1+n_{|\vec{l}_1 + \vec{l}_2|}) \,\, - \phantom{\frac{a}{b}} n_{p\eta} n_{l_1} n_{l_2} n_{|\vec{l}_1 + \vec{l}_2|} \right ] \right\}.
\ee
In the process of the derivation of this equation, we have neglected $p$ in comparison with $q_{1,2,3}$ on its RHS, denoted $\vec{l}_{1,2,3} = \vec{q}_{1,2,3}\eta$, and assumed that, on the RHS of Eq. (\ref{DS}), $n_{k\eta}$ is a much slower function of time than $h(k\eta)$. The latter fact is true because of the above-mentioned separation of scales. As a result, we can safely take out all $n$'s from the argument of the time integral on the RHS of Eq. (\ref{DS}) and substitute $D^K_0$ by $D^K$.

Note that one can reproduce $n_p$ in Eq.(\ref{nkappa}) from Eq.(\ref{kineq}) if he will use Hankel functions in place of $h(x)$, change $|A|^2 v^{i\mu}$ to $[|A_+|^2 \, v^{i\,\mu} + |A_-|^2 \, v^{-i\,\mu}]$ and put all $n$'s to zero on the RHS of Eq. (\ref{kineq}). This fact explains the physical origin of the large IR effects, which are under consideration in the present article.

\section{Solution of the kinetic equation}

The kinetic equation (\ref{kineq}) does not possess Plankian distribution as its solution, because of the violation of energy conservation in the dS background gravitational field. It can be mapped to the kinetic equation in flat space via the substitution of the harmonics $\eta^{(D-1)/2}\, h(k\eta)$ by the plane waves $e^{i \, \epsilon(k) \,t}$ in the collision integral. In the latter case, one has $\delta$ functions ensuring energy conservation on the RHS of Eq. (\ref{kineq}) instead of the integrals of $h$'s. That, in particular, means that for high-energy modes, $k\eta \gg \mu$, the kinetics in dS space is the same as in the flat one: they can scatter off each other but cannot be created in various processes involving the dS background.

Thus, suppose we have started at past infinity of the PP with some very mild density perturbation over the BD state. After the Bogolyubov rotation to the out Jost harmonics, one has some initial values of $n_p$ and $k_p$. As can be understood from the discussion in the previous paragraph, for the given $p$, the density $n_p$ and the anomalous average $k_p$ practically do not change before the moment $\eta_* \sim \mu/p$. After this moment they begin to evolve according to the coupled system of kinetic equations for $n_p$ and $k_p$. (To simplify the presentation, we do not show them here. Similar equations for the $\phi^3$ theory can be found in Refs. \cite{Akhmedov:2011pj}, \cite{Akhmedov:2012dn}.) If $k_p$ is sufficiently small, it relaxes to zero \cite{Akhmedov:2012pa}, and the problem is reduced to the solution of Eq. (\ref{kineq}).

Now, if the initial value of $n_p$ after the rotation to the out Jost harmonics is much smaller than one, we can use the following approximations:

\be
(1+n_{p\eta}) n_{l_1} n_{l_2} (1+n_{|\vec{l}_1 + \vec{l}_2|}) - n_{p\eta} (1+n_{l_1}) (1+n_{l_2}) n_{|\vec{l}_1 + \vec{l}_2|} \approx 0 \nonumber \\
(1+n_{p\eta}) n_{l_1} (1+n_{l_2}) (1+n_{|\vec{l}_1 - \vec{l}_2|}) - n_{p\eta} (1+n_{l_1}) n_{l_2} n_{|\vec{l}_1 - \vec{l}_2|} \approx n_{l_1} \nonumber \\
(1+n_{p\eta}) n_{l_1} n_{l_2} n_{|\vec{l}_1 + \vec{l}_2|} - n_{p\eta} (1+n_{l_1}) (1+n_{l_2}) (1+n_{|\vec{l}_1 + \vec{l}_2|}) \approx - n_{p\eta} \nonumber \\
(1+n_{p\eta}) (1+n_{l_1}) (1+n_{l_2}) (1+n_{|\vec{l}_1 + \vec{l}_2|}) - n_{p\eta} n_{l_1} n_{l_2} n_{|\vec{l}_1 + \vec{l}_2|} \approx 1.
\ee
Because of the rapid oscillations of $h(x)$ as $x\to\infty$, the integrals on the RHS of Eq. (\ref{kineq}) are saturated at $l_i \sim \mu$. At the same time, $p\eta \ll \mu$. Furthermore, it is natural to assume that $n_{l_i \sim 1} \ll n_{p\eta}$ in the situation of the very small initial density perturbation and the vanishing production of the high momentum modes. Hence, we can neglect the second term on the RHS of Eq. (\ref{kineq}) in comparison with the third and the fourth ones.

As a result, Eq. (\ref{kineq}) is reduced to

\be\label{solu1}
\frac{d n_{p\eta}}{d \log(p\eta)} \approx \Gamma_1 \, n_{p\eta} - \Gamma_2, \quad {\rm where} \nonumber \\
\Gamma_1 = \frac{\lambda^2 \, |A|^2}{6} \, \int \frac{d^{D-1} l_1}{(2\pi)^{D-1}} \frac{d^{D-1} l_2}{(2 \pi)^{D-1}} \int_\infty^0 dv \, v^{D-2}\times \nonumber \\ \times \R \left [v^{i\mu \phantom{\frac{a}{b}}} \, h^*(l_1) h^*(l_2) h^*\left(\left|\vec{l}_1 + \vec{l}_2\right|\right) h(l_1 v) h(l_2 v) h\left(\left|\vec{l}_1+\vec{l}_2\right| v\right) \right   ], \nonumber \\
\Gamma_2 = \frac{\lambda^2 \, |A|^2}{6} \, \int \frac{d^{D-1} l_1}{(2\pi)^{D-1}} \frac{d^{D-1} l_2}{(2 \pi)^{D-1}} \int_\infty^0 dv \, v^{D-2}\times \nonumber \\ \times \R  \left [v^{i\mu \phantom{\frac{a}{b}}} \, h(l_1) h(l_2) h\left(\left|\vec{l}_1 + \vec{l}_2\right|\right) h^*(l_1 v) h^*(l_2 v) h^*\left(\left|\vec{l}_1+\vec{l}_2\right| v\right) \right   ].
\ee
Here, $\Gamma_1$ and $\Gamma_2$ are the particle decay and production rates, correspondingly. Note that $p\eta$ is reducing to zero in the approach toward the future infinity.

The obtained equation (\ref{solu1}) has the solution with the flat stationary point distribution $n_{p\eta} = \Gamma_2/\Gamma_1$, which corresponds to the situation in which the production (gain) of particles on the level $p\eta$ is equilibrated by the particle decay (loss) from the same level. Note that here we are talking about the number density per comoving volume, which would stay constant if there were no particle decay and production processes.

The obtained solution is self-consistent for the large enough $\mu$ because then $\Gamma_2/\Gamma_1 \approx e^{-3\,\pi\mu} \ll 1$. Note that the equilibrium distribution is $n \approx e^{-3\pi\mu}$ and is not quite a Gibbons-Hawking one. Apparently, it looks like the thermal Boltzmann one, but the temperature depends on the power of the self-interaction potential. In fact, the stationary distribution in the $\phi^3$ theory is $n \approx e^{-2\pi\mu}$.

So the result of the summation of the large IR contributions may lead to the finite exact two-point functions. But what if the evolution had started with some quite strong density perturbation (which is, however, still smaller than the cosmological constant) over the BD vacuum state? Now, we are going to show that there is another very peculiar solution of the kinetic equation under consideration. See Refs. \cite{Akhmedov:2011pj}, \cite{Akhmedov:2012dn} for the similar discussion in the case of the $\phi^3$ theory.

Suppose that, due to the particle creation by the background gravitational field and by the particle decays from the other levels, the density per comoving volume on the given level with $p\eta \ll \mu$ became big in comparison with one. Taking into account the flatness of the spectrum in dS space, it is natural to expect that, for the harmonics with the low physical momenta, the density very slowly depends on its argument. Hence, we can assume that $n(p\eta) \approx n(q_{1,2,3} \eta)$ for $p\eta \ll \mu$ and $q_{1,2,3}\eta \ll \mu$.

Then, we can make the following approximations:

\be
(1+n_{p\eta}) n_{l_1} n_{l_2} (1+n_{l_3}) - n_{p\eta} (1+n_{l_1}) (1+n_{l_2}) n_{l_3} \approx 0 \nonumber \\
(1+n_{p\eta}) n_{l_1} (1+n_{l_2}) (1+n_{l_3}) - n_{p\eta} (1+n_{l_1}) n_{l_2} n_{l_3} \approx 2 \, n^3_{p\eta} \nonumber \\
(1+n_{p\eta}) n_{l_1} n_{l_2} n_{l_3} - n_{p\eta} (1+n_{l_1}) (1+n_{l_2}) (1+n_{l_3}) \approx - 2 \, n^3_{p\eta} \nonumber \\
(1+n_{p\eta}) (1+n_{l_1}) (1+n_{l_2}) (1+n_{l_3}) - n_{p\eta} n_{l_1} n_{l_2} n_{l_3} \approx 4\, n^3_{p\eta}
\ee
and accept that, on the RHS of Eq. (\ref{kineq}), the main contribution to the $l_i$ integrals comes form the region in which $l_i \ll \mu$ because  $n(x) \gg n(y)$ if $x \ll \mu$ and $y \gg \mu$. As a result, the kinetic equation reduces to

\be
\frac{d n_{p\eta}}{d\log(p\eta)} \approx - \bar{\Gamma} \, n^3_{p\eta}, \quad {\rm where} \nonumber \\
\bar{\Gamma} = \frac{\lambda^2 \, |A|^2}{3} \, \int^{|l_1|<\mu} \frac{d^{D-1} l_1}{(2\pi)^{D-1}} \int^{|l_2|<\mu} \frac{d^{D-1} l_2}{(2 \pi)^{D-1}} \int_\infty^0 dv \, v^{D-2}\, \nonumber \\
\Bigl\{3 \, |A|^2 \, A \R \left [\left(\frac{v \, l_2 \, \left|\vec{l}_1 - \vec{l}_2\right|}{l_1}\right)^{i\mu} h(l_1 v) h^*(l_2 v) h^*\left(\left|\vec{l}_1 - \vec{l}_2\right| v\right) \right] \Bigr. \nonumber \\ - \left(A^*\right)^3 \, \R \left [\left(\frac{v}{l_1 \, l_2 \, \left|\vec{l}_1 + \vec{l}_2\right|}\right)^{i\mu}\, h(l_1 v) h(l_2 v) h\left(\left|\vec{l}_1+\vec{l}_2\right| v\right) \right   ] \nonumber \\ +
\Bigl. 2 \, A^3\, \R  \left [\left(v \, l_1 \, l_2 \, \left|\vec{l}_1 + \vec{l}_2\right|\right)^{i\mu \phantom{\frac{a}{b}}} \, h^*(l_1 v) h^*(l_2 v) h^*\left(\left|\vec{l}_1+\vec{l}_2\right| v\right) \right   ] \Bigr\}.
\ee
Note that $\bar{\Gamma}$ is independent of $p$. This equation has the solution

\be\label{solu2}
n_{p\eta} \approx \frac{1}{\sqrt{2\, \bar{\Gamma} \, \log\left(\eta/\eta_{\star}\right)}},
\ee
where $\eta_{\star} = \frac{\mu}{p} \, e^{-\frac{C}{2\bar{\Gamma}}}$ and $C$ is the integration constant, which depends on the initial conditions. The obtained solution is valid if $\mu/p = \eta_* > \eta > \eta_{\star}$.

Thus, we see that there is a singular solution of the kinetic equation under consideration, which corresponds to the explosion of the particle number density per comoving volume within a finite proper time. Of course, such an explosion wins against the expansion of the PP because $D^K \propto \eta^{D-1}/\sqrt{\log{\eta/\eta_{\star}}}$. Hence, the energy-momentum tensor of the created particles becomes huge, and the backreaction has to be taken into account. As a result, the dS space gets modified. But that is the problem for a separate study. At this point, we just would like to stress that we see catastrophic IR effects even for the massive fields. It is natural to expect that, for the light fields, the situation will be even more dramatic.

\section{Comments on the contracting PP and global dS (instead of Conclusions)}

Contracting PP of dS space is interesting at least because it is complementary to the expanding PP within global dS space, and we find it quite dangerous to study such a geodesically incomplete subspace as the PP alone. Contracting PP is represented by the same metric, $ds^2 = \frac{1}{\eta}\, \left[d\eta^2 - d\vec{x}^2\right]$, as the expanding one, but now the conformal time $\eta$ flows in the proper direction --- from zero, at the past, to infinity, in the future.

From Eq. (\ref{kineq}), one can straightforwardly obtain the kinetic equation in the contracting PP, if one will consider perfectly spatially homogeneous states. The latter situation, while being stable in the expanding PP, is unstable in the contracting one under small inhomogeneous density perturbations. However, it is still instructive to consider such an ideal situation in the contracting PP\footnote{Actually it is not very hard to find the inhomogeneous extension of the kinetic equation (\ref{kineq}). In the case in which the particle density starts to depend on the spatial position $n_p = n_p(x)$, one has to substitute $\eta\, d/d\eta$ on the LHS of Eq. (\ref{kineq}) by $\eta \, \partial_\eta + \eta^2 \vec{p} \, \vec{\partial}_x$.}. In this section, we restrict ourselves to the case of the principal series.

To perform the map between the expanding and contracting PPs, both in the few-loop calculations and in the kinetic equation one just has to flip the limits of $d\eta$ integrations and change $h(x)$ and $h^*(x)$ because of the exchange between positive and negative energy states under the flip of time.

Then, it is straightforward to see that loop corrections to $D^K$ have the explicit IR divergence because particle creation starts right at the moment $\eta_0$, when we switch on self-interactions. The divergence reveals itself via the impossibility to move the $\eta_0$ to the past infinity. Because of the blueshifting of all modes, however, all the relevant kinetic processes stop after $\eta_* \sim \mu/p$. As a result, the two-loop divergence in question is proportional to $\log(\eta/\eta_0)$, when $p\eta \ll \mu$, and to $\log(\mu/p\eta_0)$, when $p\eta \gg \mu$. The prefactors are easily derivable, in view of the above discussion. Also, it is straightforward to show that for the in Jost harmonics (Bessel functions in place of $h$'s) of the contracting PP, the $k_p$ behaves similarly to that of the out Jost harmonics of the expanding PP.

In conclusion, in the contracting PP, we can find similar solutions of the kinetic equation to those that have been found in the previous section. For example, Eq. (\ref{solu1}) is mapped to

\be\label{solu4}
\frac{d n(\eta)}{d \log(\eta/\eta_0)} \approx - \Gamma_1 \, n(\eta) + \Gamma_2,
\ee
which shows a very peculiar phenomenon that, despite that in two loops $\eta_0$ --- the moment of switching on self-interactions --- cannot be taken to the past infinity, after the summation of all loops, we may find the theory at the stationary point state, $n_p = \Gamma_2/\Gamma_1$, which allows one to remove the IR cutoff, $\eta_0$.

At the same time, the solution (\ref{solu2}) is mapped to

\be\label{solu3}
n_p(\eta) \approx \frac{1}{\sqrt{2\, \bar{\Gamma} \, \log\left(\eta_{\star}/\eta\right)}},
\ee
where $\eta_{\star} = \eta_0 \, e^{C/2\bar{\Gamma}} < \mu/p$. Of course, whether the field theory state goes into Eq. (\ref{solu3}) or to Eq. (\ref{solu4}) depends on the initial conditions.

The situation in the global dS space is even more interesting. On the one hand, all sorts of contributions to the collision integral in global dS space are sums of those in the expanding and contracting PPs\footnote{For the case of the Euclidian vacuum, the relative signs of the two contributions coming from the expanding and contracting PPs can be different depending on whether $D$ is even or odd \cite{Polyakov:2012uc}. But that does not conceptually affect our discussion.}. But depending on the choice of harmonics, we either have IR divergence in $k_p$ or $k_p$ is divergent as the system advances toward the future infinity. Or an even more generic situation for $\alpha$ vacua is when $k_p$ has both types of such divergences simultaneously. As a result, there is no choice of harmonics in which $k_p$ is negligible in comparison with $n_p$, which probably means that there is no stationary state in global dS, and the backreaction on the background geometry is strong with any initial conditions.

\section{Acknowledgements}

We would like to acknowledge discussions with A.Sadofyev, Ph.Burda and A.Morozov.
This work was done under the partial financial support of the Ministry of Education and Science of Russian Federation under the contract 8207. The work of VS and FP was partially supported by the grant "Leading Scientific Schools" No. NSh-3349.2012.2. The work of VS was partially supported by the grant RFBR 11-02-01220. The work of ETA was partially supported by the grant "Leading Scientific Schools" No. NSh-6260.2010.2 and by RFBR-11-02-01227-a.

\end{document}